\documentclass[12pt]{article}

\begin{document}

\begin{center}
{\large \bf An application of the catastrophe theory to building the model of
elastic-plastic behaviour of materials.}
\par
{\large \bf Part 2. 3D model}
\par
\bigskip L. N. Maurin, I. S. Tikhomirova
\footnote{e-mail: tihomir@ivanovo.ac.ru}
\par
{\sl Physics Department, Ivanovo State University
\par
Ermaka St. 39, Ivanovo, 153025, Russia}
\end{center}

\begin{abstract}
The three-dimensional elastic-plastic deformation is considered.
The catastrophe theory underlies the construction of this process model. It
was shown that the variety of stable states consists on elastic states and
can be depicted as a lattice on T -- $\Gamma $ plane, where T is shearing
stress intensity and $\Gamma $ is shearing strain intensity.
\end{abstract}

\par
The uniaxial model of elastic-plastic deformation was built in [1]. Just as
in that article we use the catastrophe theory to construct the model. To
define the state function we introduce two variables. The first one is the
shearing stress intensity T:
\begin{center}
$T = \sqrt {\frac{1}{2} \cdot s _{ij} s _{ij}}$,
\end{center}

\par
\noindent where $s _{ij}$ are the components of stress deviator.

\noindent
The second one is the shearing strain intensity:
\begin{center}
$\Gamma = \sqrt {2 \cdot e _{ij} e _{ij} }$,
\end{center}

\par
\noindent where $e _{ij}$ are the components of strain deviator.

With these variables Hook's law takes the following form: T = G$\cdot \Gamma
$, where is the shear modulus or in dimensionless units: T = $\Gamma $.
We make a substitution here, where T stands for T/G.

To construct the model we take into account the assumption of existence of
the equilibrium static states of the deforming material. These equilibrium
states are elastic states. Plastic states are not static equilibrium. They
are implemented in the transforming from one equilibrium state (elastic
state) to another. Thus we can construct the variety of equilibrium states
that is depicted on Fig. 1a for ideal plastic materials and on Fig. 1b for
strengthening material. As it is seen from Fig. 1 the variety of stable
states is the lattice which we will call the $\Delta $-lattice, here T is
non-dimensional quantity.

So in 3D model the variety of stable states forms the lattice the same way
as in one-dimensional model. The variety of unstable states consists of
segments joining the upper end (A$_{i}$) of each rod with lower end (O$_{i+1}
$) of next rod (see the dotted lines on Fig. 2).

Just as in one-dimensional model we distinguish two types of models: the
model with parameter T and the model with parameter $\Gamma $. Different
types of transitions from one rod of $\Delta $-lattice to next rod
correspond to different types of models. Transitions in the model with
parameter T are the same as in the model with parameter $\sigma $. And
transitions in the model with parameter $\Gamma $ are the same as in the
model with parameter $\varepsilon $ (see [1] and Fig. 3). That is why we
refer the reader to [1] in oreder to omit redundant discussions.

Let us summarize the demands that the state function have to satisfy. The
state function must be smooth and has minima at each rod of $\Delta $%
-lattice and maxima for rods of additional variety. In accordance with these
demands we start to construct the state functions. Firstly we consider the
state function with parameter T. The function looks like this:

(1) \ \ \ \ \ \ \ \ \ \ \ \ \ \ \ \ \ \ \ \  $\Phi =\Phi_{n}(\Gamma \,;\,T)$,

(2a) \ \ \ \ \ \ \ \ \ \ \ \ \ \ \ $\Phi _{n}^{/}\equiv
\frac{d\Phi _{n}}{d\,\Gamma }=-\prod\limits_{s\,=\,\,-k}^{k}\,
\left( T-\varphi _{n-s}(\Gamma)\right) \,$,

(2b) \ \ \ \  $\Gamma _{i}\,\/\leq \,\Gamma \,\,<\,\,\Gamma _{i+1}$,
if $\ \Gamma _{i}$ - projection of lower end of rod onto $\Gamma $
\par
\ \ \ \ \ \ \ \ \ \ \ \ \ \ \ \ \ \ \ \ \ \ \ \ \ \
\ \ \ \ \ \ \ \ \ \ \ \ \ \ \ \ \ \ \  axis,

\ \ \ \ \ \ \ \ \ \ \ \ \ \ \ \ \ \ \ \ \ \ \ \ \ \ \ \ \
\ \ \ \ \ \ \ \ \ \ $\Gamma _{i+1}$ - projection of\ lower end of next rod
onto
\par
\ \ \ \ \ \ \ \ \ \ \ \ \ \ \ \ \ \ \ \ \ \ \ \ \ \
\ \ \ \ \ \ \ \ \ \ \ \ \ \ \ \ \ \ \  $\Gamma $ axis;

(2c) \ \ \ \ $\Gamma _{i}\,\/\leq \,\Gamma \leq \Gamma _{i+1}$,
 if $\Gamma _{i}$ - projection of lower end of rod onto $\Gamma$
\par
\ \ \ \ \ \ \ \ \ \ \ \ \ \ \ \ \ \ \ \ \ \ \ \ \ \
\ \ \ \ \ \ \ \ \ \ \ \ \ \ \ \ \ \ \  axis,

\ \ \ \ \ \ \ \ \ \ \ \ \ \ \ \ \ \ \ \ \ \ \ \ \ \ \ \ \ \
\ \ \ \ \ \  $\Gamma _{i+1}$ - projection of upper end of rod onto
$\Gamma $ axis
\par
\ \ \ \ \ \ \ \ \ \ \ \ \ \ \ \ \ \ \ \ \ \ \ \ \ \ \ \
\ \ \ \ \ \ \ \ \ \ \ \ \ \ \
 that is defined by function $\varphi _{l}$, where l$<$i;

(2d) \ \ \ \  $\Gamma _{i}\,\/<\,\,\Gamma \,\,<\,\,\Gamma _{i+1}$,
if $\Gamma _{i}$ - projection of upper end of rod onto $\Gamma $
\par
\ \ \ \ \ \ \ \ \ \ \ \ \ \ \ \ \ \ \ \ \ \ \ \ \ \
\ \ \ \ \ \ \ \ \ \ \ \ \ \ \ \ \ \ \  axis,

\ \ \ \ \ \ \ \ \ \ \ \ \ \ \ \ \ \ \ \ \ \ \ \ \ \ \ \ \ \
\ \ \ \ \ \  $\Gamma _{i+1}$ - projection of lower end of next rod onto
\par
\ \ \ \ \ \ \ \ \ \ \ \ \ \ \ \ \ \ \ \ \ \ \ \ \ \
\ \ \ \ \ \ \ \ \ \ \ \ \ \ \ \ \ \ \  $\Gamma $ axis;

(2e) \ \ \ \  $\Gamma _{i}\,\/<\,\,\Gamma \,\,\leq \,\,\Gamma _{i+1}$,
if $\Gamma _{i}$ - projection of upper end of rod onto $\Gamma $
\par
\ \ \ \ \ \ \ \ \ \ \ \ \ \ \ \ \ \ \ \ \ \ \ \ \ \
\ \ \ \ \ \ \ \ \ \ \ \ \ \ \ \ \ \ \  axis,

\ \ \ \ \ \ \ \ \ \ \ \ \ \ \ \ \ \ \ \ \ \ \ \ \ \ \ \ \ \ \
\ \ \ \ \  $\Gamma _{i+1}$ - projection of upper end of next rod
onto
\par
\ \ \ \ \ \ \ \ \ \ \ \ \ \ \ \ \ \ \ \ \ \ \ \ \ \
\ \ \ \ \ \ \ \ \ \ \ \ \ \ \ \ \ \ \  $\Gamma $ axis;

(3) \ \ \ \  $\Phi _{n}(\Gamma _{n+1})=\Phi
_{n+1}(\Gamma _{n+1})$ - lacing condition, where $n<(N-1)$,
\par
\noindent N - number of intervals on $\Gamma$ axis.

\noindent
Here for rods of $\Delta $-lattice

(4) \ \ \ \ \ \ \ \ \ \ \
\ \ \ \ \ \ \ \ \ \ \ \ $i=2m$, \ \ \ \ \ \ $\varphi _{i}(\Gamma )=\Gamma
\,\,-\,\sum\limits_{l=0}^{\frac{i}{2}-1}\Delta _{l},$
\par
\noindent and for rods of additional variety

(5) \ \ \ \ \ \ \ \ \ \ \ \ \ \ \ \ \
$i=2\cdot m+1$, \ \ \ \ \  $\varphi _{i}(\Gamma )=\,\,k_{i}\cdot \left(
\Gamma \,\,-\,\sum\limits_{l=0}^{\frac{i-1}{2}}\Delta _{l}\right)$,

\par
\noindent
where $k_{i}=\frac{\left( \Gamma -\sum\limits_{i=0}^{m-1}\Delta _{l}\right)
}{\left( \Gamma -\sum\limits_{i=0}^{m}\Delta _{l}\right) }=1+\frac{\Delta
_{m}}{\Gamma -\sum\limits_{i=0}^{m}\Delta _{l}}$, and $\Gamma $ is equal to
the strain in point A$_{m}$.

As it is seen from equations (1) - (3) the state function is constructed of
all $\,\varphi _{i}\left( \Gamma \right) $ existing on the examining part of
the strain axis $\Gamma $. It is not difficult to test (using (1) - (5))
that the state function $\Phi _{n}$ is minimum on all rods of $\Delta$
-lattice (i. e. these states are stable) and maximum on all rods of
additional variety (unstable states). It is necessary to note that the end
points of rods are degenerated critical points.

The second state function type is the state function with parameter $\Gamma$.
We define this function as:

(6) \ \ \ \ \ \ \ \ \ \ \ \ \ \ \
$\Phi \,\,=\,\,\Phi _{n}\left( T\,;\,\,\Gamma \right) $, \ \ \ \ \
$\Phi _{n}(T_{n+1})\,=\,\Phi _{n+1}(T_{n+1})$,

(7) \ \ \ \ \ \ \ \ \ \  $\Phi _n^ / = \frac{d\Phi _n }{d\,T} = -\prod\limits_{i = 2n}^
{2(N - 1)} {\left( {\Gamma - \psi _i (T)} \right)} = \prod\limits_{i = 2n}^
{2(N - 1)} {\left( {\psi _i (T) - \Gamma } \right)} $,

\par
\noindent
defined for region $T_{n} <T \leq T_{n+1} $, if $n \neq 0$, and for
$T_{n} \leq T \leq T_{n+1}$, if $\ n = 0$,
here N stands for number of rods of $\Delta$-lattice.

Here $\psi _{i}(T)=T + \sum\limits_{l=0}^{\frac{i}{2}-1}\Delta _{l}$,
for the rods of $\Delta $-lattice, where $i=2m$,
\par
\noindent
and for the rods of additional variety:
\par
\begin{center}
$\psi _{i} (T) = \frac{1}{k_{i} } \cdot
T + \sum\limits_{l=0}^{\frac{i-1}{2}}\Delta _{l} $, where $i=2\cdot m+1$.
\end{center}
\par
\noindent
Here k$_{i}$ is the same as in the case of a model with parameter T.

Like the previous state function (with parameter T) we use all of functions
$\psi _{i}\left( T\right) $ existing on the examining part of the stress axis
T to construct this function, i. e. it is formed by means of all rods of
$\Delta $-lattice and additional variety existing on the examining part of
the T-axis. It is easy to verify that the state function $\Phi _{n}\left(
T\,; \Gamma \right)$ (see (6) - (7)) also satisfies the above demands.

So the 3D model of elastic-plastic deformation was built both with parameter
$\Gamma$ and parameter T.

\small
\bigskip
\begin{center}
{\bf References}
\end{center}
\medskip
\small
\begin{itemize}
\item[[1{]}]
An application of the catastrophe theory to building the model of
elastic-plastic behaviour of materials. Part 1. Uniaxial deformation (stress)
//

cond-mat/0111304 (http://xxx.lanl.gov/abs/cond-mat/0111304)
\end{itemize}
\par
\vspace{10mm}
\textbf{Fig. 1 a, b.} The $\Delta $-lattice of ideal plastic (a) and strengthening (b)
materials.

\textbf{Fig. 2 a, b.} The $\Delta $-lattice and additional variety of ideal plastic (a)
and strengthening (b) materials.

\textbf{Fig. 3 a, b.} The possible transitions in the model with parameter T (a) and in
the model with parameter $\Gamma $ (b).

\end{document}